\documentstyle[prl,aps,amsfonts,amssymb,twocolumn,psfig,rotate]{revtex}
\begin{document}
\def\u{\bbox}
\def\mathcal#1{{\cal #1}}
\def\phi{\varphi}
\def\epsilon{\varepsilon}
\def\Bbb{\relax}

\draft
\title{Critical Attractor and Universality in a  
Renormalization Scheme for Three Frequency Hamiltonian Systems}
\author{C. Chandre and H.R. Jauslin}
\address{Laboratoire de Physique, CNRS, Universit\'e de Bourgogne,
BP 400, F-21011 Dijon, France}
\maketitle

\begin{abstract}
We study an approximate renormalization-group transformation to analyze
the breakup of invariant tori for three degrees of freedom Hamiltonian systems.
The scheme is implemented for the spiral mean torus.
We find numerically that the critical surface is the stable manifold
of a critical nonperiodic attractor. We compute scaling exponents associated
with this fixed set, and find that they can be expected to be universal.
\end{abstract}


The breakup of invariant tori is one of the key mechanism of the transition to 
chaos in Hamiltonian dynamics.
For two dimensional systems and for frequencies like
the golden mean, it has been observed that, at the transition, a
sequence of periodic orbits approaches geometrically a torus of the
given frequency, with
a nontrivial scaling behavior~\cite{greene,kadanoff,shenkerkadanoff}. 
This self-similarity has been described in terms of a nontrivial fixed point 
of a renormalization-group transformation
\cite{mackay,govin,chandre,cgjk,abad}.
The sequence of periodic orbits responsible for the breakup is generated 
by the continued fraction expansion of the frequency.
For the extension to 
systems with three degrees of freedom (d.f.) involving three incommensurate
frequencies,
we lack a theory that generalizes the continued fractions.
Numerically, three d.f.\ Hamiltonian systems (or equivalently four dimensional 
volume-preserving maps) have been studied with an extension of 
Greene's criterion
\cite{maohelleman,artusocasati,tompaidis,kurosaki}.
The conclusion of these analysis was that there is no
geometrical accumulation of periodic orbits
around the critical torus, and thus absence of universality.\\
The aim of this Letter is to show that one can still expect {\em universal}
behavior in the breakup of invariant tori, described by a critical  
attractor of a renormalization-group transformation. The idea is that all
Hamiltonians attracted by renormalization to this set will display 
sequences of scaling factors that appear in a different order 
but with a universal statistical
distribution. The dominant unstable Lyapunov exponent characterizes the
approach to criticality of the universality class.\\
Nonperiodic attractors for renormalization maps have been conjectured 
and observed in statistical mechanics~\cite{mckay,eckmann} and 
dynamical systems~\cite{lanford,rand}.
In Ref.~\cite{satija1}, a strange attractor was found for renormalization
of circle maps, and in Ref.~\cite{satija2} similar evidence was found
for area-preserving maps in two dimensions, from the scaling analysis
of periodic orbits. The origin of randomness
in these latter studies is due to the randomness of 
the sequence of continued fraction 
approximants for an ensemble of the considered frequencies.
In contrast, in the present three d.f.\ case,
the rational approximants are a regular sequence, obtained
by iteration of a {\em single} unimodular matrix, which allows us to define a 
renormalization transformation with a {\em fixed} frequency vector.
The idea is to set up a transformation $\mathcal{R}$ that maps a 
Hamiltonian $H$ into a rescaled Hamiltonian $\mathcal{R}(H)$, 
such that irrelevant degrees of freedom are eliminated.
The transformation $\mathcal{R}$ (which will be defined precisely below)
should have roughly the following properties:
$\mathcal{R}$ has an attractive integrable fixed point $H_0$
(trivial fixed point) that has a smooth
invariant torus of a given frequency ${\u \omega}_0$. Every Hamiltonian in its
domain of attraction $\mathcal{D}$ has 
a smooth invariant torus with frequency vector ${\u \omega}_0$. 
The aim is to show that there
is another fixed set $\Lambda$ which lies on the boundary $\partial
\mathcal{D}$ (the
critical surface) and that is attractive for every Hamiltonian on 
$\partial \mathcal{D}$.
The numerical implementation for two d.f.~\cite{govin,chandre,cgjk} gives
support to this picture. This is by no means trivial since the construction
of the renormalization iteration is based on properties that are valid
close to $H_0$. The numerical results indicate that the domain of
convergence of the iteration $\mathcal{R}$ indeed extends up 
to the critical surface. A mathematical justification
of this observation, and the formulation of conditions for its validity,
are completely open problems.\\
In this paper, we study the extension of these ideas to 
systems with three frequencies, 
by analyzing an approximate renormalization transformation
based on the work of Escande and 
Doveil \cite{escandedoveil,escande,benfatto,mackaymeiss}. Its properties 
are found in agreement with the general picture described above.
We find 
universal exponents associated with the fixed set $\Lambda$. They are
universal in the sense that they
only depend on the frequency vector considered, and not on the chosen
one-parameter family.
These results give new insights for the set-up of 
an exact renormalization scheme in the spirit of Refs.\
\cite{govin,chandre,cgjk,abad}.\\
The transformation we define acts on the following class of Hamiltonians 
with three d.f., 
quadratic in the actions $\u{A}=(A_1,A_2,A_3)$, and described by three
even scalar functions of the angles $\u{\phi}=(\phi_1,\phi_2,\phi_3)$:
\begin{eqnarray}
  H(\u{A},\u{\phi})=&&\frac{1}{2}\left(1+m(\u{\phi})\right)
        (\u{\Omega}\cdot\u{A})^2 \nonumber \\
     &&+\left[\u{\omega}_0+g(\u{\phi})\u{\Omega}
	\right]\cdot\u{A}+f(\u{\phi}),\label{hamiltonian}
\end{eqnarray}
where $m$, $g$, and $f$ are of zero average.
The vector $\u{\omega}_0$ is the frequency vector of the considered torus and
$\u{\Omega}$ is a vector not parallel to $\u{\omega}_0$, with
norm one: $\Vert {\u \Omega}\Vert=\left( |\Omega_1|^2+|\Omega_2|^2+
|\Omega_3|^2\right)^{1/2}=1$.
This model is the simplest one involving invariant tori with three frequencies.
It can be thought as an intermediate case between two and three d.f.
Its behavior is more complex than in two d.f.: For instance, the
geometric Aubry-Mather theory is, to our knowledge, 
not available~\cite{moser}.\\ 
The transformation $\mathcal{R}$ is defined for a fixed frequency vector
${\u \omega}_0$ with
three incommensurate components. 
We choose
$
{\u \omega}_0=(\sigma^2,\sigma,1),
$
where $\sigma\approx 1.3247$ is the spiral mean: it satisfies 
$\sigma^3=\sigma+1$ \cite{kimostlund,bollt}.
Since ${\u \omega}_0$ is Diophantine,
the KAM theorem applies to Hamiltonians (\ref{hamiltonian}) (although they
are isoenergetically degenerate \cite{chandrejauslin}) and
shows the existence of a torus $\mathcal{T}$ with frequency vector 
${\u \omega}_0$ for a sufficiently small and smooth perturbation consisting
of $m$, $g$ and $f$. The
domain of existence of  $\mathcal{T}$ corresponds to a neighborhood 
of the trivial fixed point 
\begin{equation}
\label{eqn:h0}
H_0({\u A})=({\u \Omega}\cdot{\u A})^2/2+{\u \omega}_0\cdot{\u A},
\end{equation}
for which  $\mathcal{T}$ is located at ${\u A}=0$.
From some of its properties, $\sigma$ plays a similar role
as the golden mean in the two d.f.\ case~\cite{kimostlund}. 
The analogy comes from the fact that 
one can generate rational approximants by iterating a {\em single}
unimodular matrix $N$.
In what follows, we denote {\em resonance} an element of the sequence 
$\{ {\u \nu}_k=N^{k-1}{\u \nu}_1, k\geq 1\}$ where ${\u \nu}_1=(1,0,0)$ and
$$
N=\left(\begin{array}{ccc} 0 & 0 & 1 \\ 
                             1 & 0 & 0\\
			     0 & 1 & -1
            \end{array}\right).
$$
The word resonance refers to the fact that the small denominators 
${\u \omega}_0\cdot {\u \nu}_k$ that appear in the perturbation series or
in the KAM iteration, tend to zero geometrically as $k$ increases
($
{\u \omega}_0 \cdot {\u \nu}_k = \sigma^{2-k} \to 0 \mbox{ as } k\to \infty
$).
Our hypothesis (which is also the starting point of a generalization
of Greene's criterion in Ref.\ \cite{artusocasati}) is that this sequence 
plays a leading role in the breakup of the invariant torus 
with frequency vector
${\u \omega}_0$. We build an approximate scheme by considering the three
main resonances ${\u \nu}_1$, ${\u \nu}_2$, and ${\u \nu}_3$. The 
renormalization focuses on the next smaller scale represented by the 
resonances ${\u \nu}_2$, ${\u \nu}_3$, together with
${\u \nu}_4=N{\u \nu}_3= {\u \nu}_1-{\u \nu}_3$. 
It includes a partial elimination
of the perturbation (the part which can be considered
nonresonant on the smaller
scale, namely the mode ${\u \nu}_1$), a shift of the resonances, a rescaling
of the actions and of the energy, and a translation in the action variables.
It is, in spirit, close to the type of transformation considered in
Refs.\ \cite{cgjk,abad}.\\
The approximations involved in this scheme are the two main ones used by Escande
and Doveil:\\
a] A quadratic approximation in the actions: the rescaled Hamiltonian
$\mathcal{R}(H)$ is in general, higher than quadratic in the actions;
in order to remain in the same family of Hamiltonians (\ref{hamiltonian}),
we neglect these higher order terms.\\
b] A three-resonance approximation: we only keep the three main resonances
at each iteration of the transformation. This is the simplest generalization
to three d.f.\ of the Escande-Doveil approach.

\paragraph*{Renormalization transformation.---}

The approximate transformation we define acts on a reduced family of Hamiltonians
(\ref{hamiltonian}). We consider the three most relevant Fourier modes
$\{ {\u \nu}_i, \, i=1,2,3\}$ at each step of the transformation.
Hamiltonian (\ref{hamiltonian}) can be written as
$$
H({\u A},{\u \varphi})=H_0({\u A})+\sum_{i=1}^3 h_i({\u A})\cos({\u \nu}_i
\cdot {\u \varphi}),
$$
where $H_0$ is given by Eq.\ (\ref{eqn:h0}), and
$
h_i({\u A})=m_{{\nu}_i}({\u \Omega}\cdot {\u A})^2/2
+g_{{\nu}_i}{\u \Omega}\cdot {\u A}+f_{{\nu}_i}.
$\\
Our transformation combines thus five steps:\\
1] A canonical transformation that eliminates the first main resonance
${\u \nu}_1$ to
the order $O(\varepsilon)$.
This is performed by a Lie transformation $\mathcal{U}_S:({\u \phi},{\u A})
\mapsto ({\u \phi}',{\u A}')$,
generated by 
$
S({\u A},{\u \phi})=S_1({\u A})\sin({\u \nu}_1\cdot{\u\phi})
$.
The Hamiltonian expressed in the new coordinates is given by
$
H'=\exp(\hat{S})H \equiv H+\{S,H\}+\{S,\{S,H\}\}/2!+\cdots,
$
where $\{ \, , \, \}$ is the Poisson bracket between two scalar
functions of the actions and angles.
The generating function $S$ is determined by the requirement that the order
$O(\epsilon)$ of the mode ${\u \nu}_1$ vanishes:
$
\{S,H_0\}+h_1({\u A})\cos({\u \nu}_1\cdot{\u \phi})=0.
$
This equation has the solution
$
S({\u A},{\u \phi})=-h_1({\u A})\sin({\u \nu}_1\cdot{\u \phi})
/{\u \omega}({\u A})\cdot {\u \nu}_1,
$
where ${\u \omega}({\u A})={\u \omega}_0+({\u \Omega}\cdot{\u A}){\u \Omega}$.
This step generates arbitrary orders in the action variables. In order to map
the family of Hamiltonians (\ref{hamiltonian}) into itself, we expand $H'$ to
quadratic order in the actions, and we neglect higher orders. The
justification for this approximation is that as  the torus is located
at ${\u A}=0$ for $H_0$, one can expect that for small $\varepsilon$,
it is close to ${\u A}=0$. We notice that $h_2({\u A})$ and $h_3({\u A})$
are not changed to the order $O(\varepsilon^3)$.
Furthermore, we neglect all the Fourier modes except ${\u 0}$, ${\u \nu}_2$,
${\u \nu}_3$, and ${\u \nu}_4$. We expand the Hamiltonian to the
order $O(\varepsilon^3)$. This leads to the expression of $H'$:
\begin{eqnarray}
H'=&&H_0+h_2\cos({\u \nu}_2\cdot{\u \varphi})
+h_3\cos({\u \nu}_3\cdot{\u \varphi}) \nonumber \\
&&+\langle \{S,h_1\}\rangle/2 + \{ S,h_3\}, \label{eqn:H'}
\end{eqnarray}
where $\langle \, \rangle$ denotes the mean value defined as
$
\langle h \rangle ({\u A}) = \int_{{\Bbb T}^3} \,
h({\u A},{\u \varphi})\, d^3 {\u \varphi}/(2\pi)^3.
$
The last term of Eq.\ (\ref{eqn:H'}) contains
the Fourier mode ${\u \nu}_4$ of amplitude 
\begin{equation}
\label{eqn:h4}
h_4({\u A})=(S_1{\u \nu_1}\cdot {\u \partial} h_3 +h_3 {\u \nu}_3\cdot 
{\u \partial} S_1)/2,
\end{equation}
where ${\u \partial}$ denotes the derivative with respect to ${\u A}$.
We expand $h_4$ to quadratic order in the actions.\\
2] A shift of the resonances ${\u \nu}_k\mapsto {\u \nu}_{k-1}$: 
a linear transformation 
$(\u{A},\u{\varphi})\mapsto (N^{-1}\u{A},\tilde{N}\u{\varphi})$, 
where $\tilde{N}$ is $N$ transposed.
This step changes 
the frequency ${\u \omega}_0$ into $\tilde{N}{\u \omega}_0=
\sigma^{-1}{\u \omega}_0$ (since ${\u \omega}_0$ is an eigenvector of 
$\tilde{N}$ by construction).\\
3] We rescale the energy (or equivalently the time)
by a factor $\sigma$, in order to keep
the frequency fixed at ${\u \omega}_0$.
The vector ${\u \Omega}$
is changed into $\tilde{N}{\u \Omega}$. We define the image ${\u \Omega}'$
of ${\u \Omega}$ by ${\u \Omega}'=\tilde{N}{\u \Omega}/\Vert 
\tilde{N}{\u \Omega}\Vert$, in order to have ${\u \Omega}'$ of unit norm.
We rewrite the mean-value term of Eq.\ (\ref{eqn:H'}) as
$
\langle \{S,h_1\}\rangle =\mu({\u \Omega}\cdot {\u A})^2+
a{\u \Omega}\cdot {\u A} + \mbox{const}.
$\\
4] In order to keep the image of $g$ with zero mean-value, we eliminate $a$
by a translation in the action variables ${\u A}\mapsto{\u A}+{\u a}$ (of order
$O(\varepsilon^2)$ in the ${\u \Omega}$ direction).
The constant part of the quadratic term of the resulting Hamiltonian is
$\sigma\Vert\tilde{N}{\u \Omega}\Vert^2(1+\mu)({\u \Omega}'\cdot
{\u A})^2/2$.\\
5] In order to map this Hamiltonian back into the form
(\ref{hamiltonian}),
we rescale the actions:
$
\hat{H}({\u A},{\u \varphi})=
\lambda H\left({\u A}/\lambda,{\u \varphi}\right),
$
with $\lambda=\sigma\Vert \tilde{N} {\u \Omega}\Vert ^2 (1+\mu)$.\\
In summary, the transformation is equivalent to a mapping acting on a 
11-dimensional space 
\begin{eqnarray*}
&&(m_{\nu_1},g_{\nu_1},f_{\nu_1},m_{\nu_2},g_{\nu_2},f_{\nu_2},
m_{\nu_3},g_{\nu_3},f_{\nu_3},{\u \Omega}) 
\mapsto \\
&& \qquad \qquad (m_{\nu_1}',g_{\nu_1}',f_{\nu_1}',m_{\nu_2}',g_{\nu_2}',
f_{\nu_2}',m_{\nu_3}',g_{\nu_3}',f_{\nu_3}',{\u \Omega}'),
\end{eqnarray*}
defined by the following relations
\begin{eqnarray}
&& m_{\nu_i}'=m_{\nu_{i+1}}/(1+\mu),\nonumber  \\
&& g_{\nu_i}'=\sigma\Vert\tilde{N}{\u\Omega}\Vert g_{\nu_{i+1}},\nonumber\\
&& f_{\nu_i}'=\lambda\sigma f_{\nu_{i+1}}, 
\qquad \mbox{ for } i=1,2 \nonumber\\
&& m_{\nu_3}'=2h_4^{(2)}/(1+\mu), \nonumber\\
&& g_{\nu_3}'=\sigma\Vert\tilde{N}{\u\Omega}\Vert h_4^{(1)},\nonumber\\
&& f_{\nu_3}'=\lambda\sigma h_4^{(0)},\nonumber\\
&& {\u \Omega}'=\tilde{N}{\u \Omega}/ \Vert \tilde{N}{\u \Omega} \Vert,
\label{eqn:map}
\end{eqnarray}
where $h_4^{(i)}$ is the coefficient in $(\u{\Omega}\cdot\u{A})^i$ of $h_4$
given by Eq.\ (\ref{eqn:h4}).

\paragraph*{Critical surface of the transformation.---}

The numerical implementation of this scheme shows that there are
two main domains separated by a {\em critical surface}:
one where the iteration converges to $H_0$ and the other
where it diverges to infinity.
The renormalization-group picture for two-dimensional
systems with golden mean frequency showed that this surface is 
the stable manifold of a nontrivial fixed
point (or nontrivial fixed set related to this nontrivial fixed point
by symmetries \cite{chandre}). Here, we cannot expect any relatively
stable nontrivial fixed point: this has been highlighted in 
Refs.\ \cite{artusocasati,mackaymeiss} and can
be explained by considering the map (\ref{eqn:map}).
This map has no stable point (there is only
one hyperbolic fixed point which corresponds to ${\u \omega}_0$).
The eigenvalues of $\tilde{N}$ are $\sigma^{-1}$ and
$-\sqrt{\sigma}e^{\pm i\alpha}$ where 
$\alpha\approx 2\pi \times 0.1120$.
The map (\ref{eqn:map}) leads asymptotically to a rotation of angle $\alpha$.
As $\alpha$ is close to $2\pi/9$, we expect the results to oscillate 
{\em approximately} with period 9 as it has been observed in 
Ref.\ \cite{artusocasati}. 
Figure 1 shows the scaling factor $\lambda_{k+9}$
as a function of $\lambda_k$ after $k$ iterations on the critical
surface. The points near the
diagonal correspond to approximate period 9 behavior. There are however
strong deviations from this behavior.
Figure 2 shows the statistical distribution of the scaling factors.\\
The iterations on the critical surface converge
to a nonperiodic bounded set $\Lambda$. Figure 3
shows the projection of $\Lambda$ on the plane 
$(g_{\nu_1},m_{\nu_1})$. This set
has a codimension 1 stable manifold, i.\ e., one expansive
direction transverse to the critical surface.
This set plays, for the system we consider, the same role as the nontrivial
fixed point of the renormalization-group transformation for quadratic
irrational frequencies in two d.f.\ Hamiltonian systems.
In particular, its existence implies 
{\em universality} for one-parameter families crossing the critical surface.
We define exponents that characterize the universality class
associated with the spiral mean.
The mean-rescaling is defined by
$
\lambda=\lim_{n\to \infty} \left( \prod_{k=1}^{n} \lambda_k \right)^{1/n},
$
where $\lambda_k$ is the value of the rescaling after $k$ iterations on the
critical surface. We also calculate the largest Lyapunov exponent 
$\kappa$.
The result we found is
that these limits do not depend on the point on the critical surface where
we start the iteration nor on the initial choice of ${\u \Omega}$.
The coefficients $\kappa$ and $\lambda$ depend only on ${\u
\omega}_0$. Numerically, we find $\kappa \approx 0.5870$ and 
$\lambda \approx 2.6640$.


In conclusion, the numerical results indicate
that for the spiral mean, the critical surface 
of the approximate renormalization transformation
is the stable manifold of a critical set 
instead of a nontrivial fixed point.
This feature depends strongly on the characteristics of the 
eigenvalues of $N$, and consequently we do not exclude a priori that some
specific choices of the frequency vector (or equivalently of $N$) could
lead to a nontrivial fixed point of a renormalization-group transformation.

We acknowledge
useful discussions with G.\ Benfatto, A.\ Celletti, G.\ Gallavotti, H.\ Koch,
J.\ Laskar, and R.S.\ MacKay.
Support from EC Contract No.\ ERBCHRXCT94-0460 for the project
``Stability and Universality in Classical Mechanics'' is acknowledged.

\newpage
\begin{figure}
\unitlength=1cm
\begin{center}
\begin{picture}(6,6)
\put(0,0){\psfig{figure=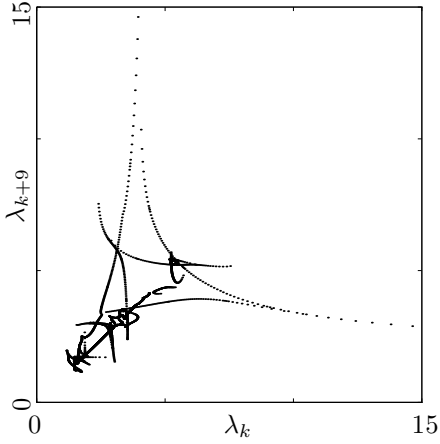,height=6cm,width=6cm}}
\put(2.7,-0.2){$\lambda_k$}
\put(-0.2,2.6){\rotate{$\lambda_{k+9}$}}
\put(-0.1,0.1){\rotate{0}}
\put(-0.1,5.2){\rotate{15}}
\put(0.1,-0.2){0}
\put(5.2,-0.2){15}
\end{picture}
\end{center}
\caption{$\lambda_{k+9}$ as a function of $\lambda_k$.}
\end{figure}

\begin{figure}
\unitlength=1cm
\begin{center}
\begin{picture}(6,6)
\put(0,0){\psfig{figure=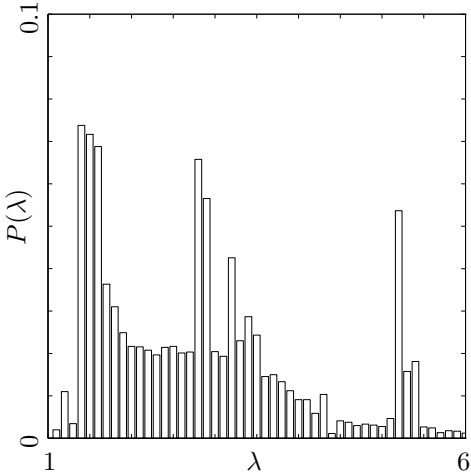,height=6cm,width=6cm}}
\put(3,-0.2){$\lambda$}
\put(-0.2,2.8){\rotate{$P(\lambda)$}}
\put(0.3,-0.2){1}
\put(5.8,-0.2){6}
\put(0,0.1){\rotate{0}}
\put(0,5.6){\rotate{0.1}}
\end{picture}
\end{center}
\caption{Distribution of the values of the rescalings $\lambda$
of the critical attractor $\Lambda$.}
\end{figure}

\begin{figure}
\unitlength=1cm
\begin{center}
\begin{picture}(6,6)
\put(0,0){\psfig{figure=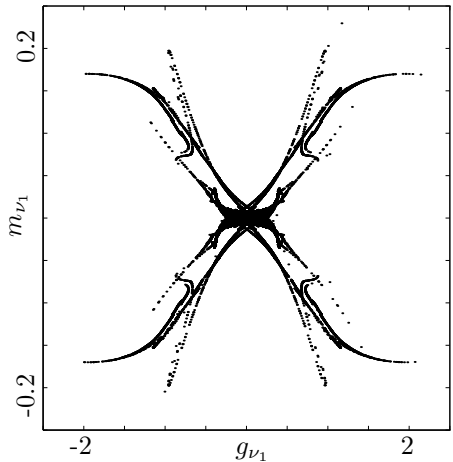,height=6cm,width=6cm}}
\put(3,-0.1){$g_{\nu_1}$}
\put(0,2.8){\rotate{$m_{\nu_1}$}}
\put(0.8,-0.1){-2}
\put(5.2,-0.1){2}
\put(0.1,0.4){\rotate{-0.2}}
\put(0.1,5.1){\rotate{0.2}}
\end{picture}
\end{center}
\caption{Projection on the plane $(g_{\nu_1},m_{\nu_1})$
of the critical attractor $\Lambda$}
\end{figure}

\newpage


\begin{references}

\bibitem{greene}
J.M.~Greene, J.\ Math.\ Phys.\ (N.Y.) {\bf 20}, 1183 (1979).

\bibitem{kadanoff}
L.P.\ Kadanoff, Phys.\ Rev.\ Lett.\ {\bf 47}, 1641 (1981).

\bibitem{shenkerkadanoff}
S.J.\ Shenker and L.P.\ Kadanoff, J.\ Stat.\ Phys.\ {\bf 27}, 631 (1982).

\bibitem{mackay}
R.S.\ MacKay, Physica (Amsterdam) {\bf 7D}, 283 (1983).

\bibitem{govin}
M.~Govin, C.~Chandre, and H.R.~Jauslin, Phys.\ Rev.\ Lett.\ {\bf 79}, 3881 (1997).

\bibitem{chandre}
C.~Chandre, M.~Govin, and H.R.~Jauslin, Phys.\ Rev.\ E {\bf 57}, 1536 (1998).

\bibitem{cgjk}
C.~Chandre, M.~Govin, H.R.~Jauslin, and H.~Koch, Phys.\ Rev.\ E {\bf 57},
6612 (1998).

\bibitem{abad}
J.J.~Abad, H.~Koch, and P.~Wittwer, Nonlinearity {\bf 11}, 1185 (1998).

\bibitem{maohelleman}
J.M.~Mao and R.H.G.~Helleman, Nuovo Cimento {\bf 104}B, 177 (1989).

\bibitem{artusocasati}
R.~Artuso, G.~Casati, and D.L.~Shepelyansky, Europhys.\ Lett.\ {\bf 15},
381 (1991).

\bibitem{tompaidis}
S.\ Tompaidis, Experimental Mathematics {\bf 5}, 197 (1996).

\bibitem{kurosaki}
S.~Kurosaki and Y.\ Aizawa, Prog.\ Theor.\ Phys.\ {\bf 98}, 783 (1997).

\bibitem{mckay}
S.R.~McKay, A.N.~Berker, and S.~Kirkpatrick, Phys.\ Rev.\ Lett.\ {\bf 48},
767 (1982).

\bibitem{eckmann}
B.~Derrida, J.P.~Eckmann, and A.~Erzan, J.\ Phys.\ A: Math.\ Gen.\ {\bf 16},
893 (1983).

\bibitem{lanford}
O.E.\ Lanford, in {\em Statistical Mechanics and Field Theory:
Mathematical Aspects}, edited by T.C.\ Dorlas, N.M.\ Hugenholtz,
and M.\ Winnink (Springer-Verlag, Berlin,1986).

\bibitem{rand}
D.A.\ Rand, Proc.\ R.\ Soc.\ Lond.\ A {\bf 413}, 45 (1987).

\bibitem{satija1}
D.K.~Umberger, J.D.~Farmer, and I.I.~Satija, Phys.\ Lett.\ A {\bf 114}, 341
(1986).

\bibitem{satija2}
I.I.~Satija, Phys.\ Rev.\ Lett.\ {\bf 58}, 623 (1987).

\bibitem{escandedoveil}
D.F.\ Escande and F.~Doveil, J.\ Stat.\ Phys.\ {\bf 26}, 257 (1981).

\bibitem{escande}
D.F.\ Escande, Phys.\ Rep.\ {\bf 121}, 165 (1985).

\bibitem{benfatto}
C.~Chandre, H.R.~Jauslin, and G.~Benfatto, J.\ Stat.\ Phys.\ {\bf 94},
to appear (1999).

\bibitem{mackaymeiss}
R.S.\ MacKay, J.D.\ Meiss, and J.~Stark, Phys.\ Lett.\ A {\bf 190}, 417 (1994).

\bibitem{moser}
J.\ Moser, Ergod.\ Th.\ \& Dynam.\ Sys.\ {\bf 8}, 251 (1988).

\bibitem{kimostlund}
S.\ Kim and S.\ Ostlund, Phys.\ Rev.\ A {\bf 34}, 3426 (1986).

\bibitem{bollt}
E.M.~Bollt and J.D.~Meiss, Physica (Amsterdam) {\bf 66D}, 282 (1993).

\bibitem{chandrejauslin}
C.~Chandre and H.R.~Jauslin, J.\ Math.\ Phys.\ {\bf 39}, 5856 (1998).

\end{references}
\end{document}